# Local decomposition at the ferroelectric-antiferroelectric interphase boundaries and formation of mesoscopic structures in lead lanthanum titanate-zirconate solid solutions.

V.M. Ishchuk[1], Z.A. Samoylenko[2], and V.L. Sobolev[3]

[1] *Institute for Single Crystals, National Academy of Sciences of Ukraine, 61001 Kharkov, Ukraine*

[2] *Donetsk Physical-Engineering Institute, National Academy of Sciences of Ukraine, 81114 Donetsk, Ukraine*

[3] *Department of Physics, South Dakota School of Mines and Technology, Rapid City, SD 57701*

E-mail: vladimir.sobolev@sdsmt.edu

**Abstract**

The results of the first investigations of the process of local decomposition of lead-lanthanum zirconate-titanate solid solution in the vicinity of the interphase boundaries separating domains of coexisting ferroelectric and antiferroelectric phases are presented. The solid solution with the composition: 6 at% of La, 73 at% of Zr, and 27 at% of Ti is considered. The kinetics of the local decomposition of this solid solution in the process of aging of samples quenched down to room temperature from the paraelectric phase is studied. The mechanisms that define the kinetics of the process of establishment of the equilibrium state of coexisting domains of ferroelectric and antiferroelectric phases are discussed.

# 1. Introduction

Compounds with inhomogeneous state of coexisting phases attract a great deal of attention recently. The reason for such interest is caused by the fundamental problems of multiphase system stability and phase transitions between phases on the one hand, and by the fact that these materials manifest extremely high values of physical parameters on the other hand. The last circumstance makes these substances very promising for a variety of applications. In fact several such compounds are already actively used in different types of devices.

Lead zirconate titanate based solid solutions (PZT: $PbZrO_3 - PbTiO_3$), lead magnesium niobate titanate (PMN-PT: $Pb(Mg_{1/3}Nb_{2/3})O_3 - PbTiO_3$), and lead zinc niobate – lead titanate (PZN-PT: $Pb(Zn_{1/3}Nb_{2/3})O_3 - PbTiO_3$) are several ferroelectric (FE) materials that are the examples of such substances with inhomogeneous states of coexisting phases. In all these compounds the unusually high values of piezoelectric parameters have been observed in precisely those solid solutions which belong to the so-called morphotropic region on the phase states diagram. This morphotropic region of the phase diagram is characterized by the coexistence of domains of ferroelectric (FE) phases with tetragonal and rhombohedral types of crystal cell distortions [1-5].

Situations when compounds with inhomogeneous states of coexisting phases are important for technological applications are widely known among magnetic and superconducting materials. Here we mention a special group of magnetic substances, namely, hard magnetic materials. It has been found, that to store the high value of magnetic energy these systems have to be essentially inhomogeneous [6-7].

The type II superconductors comprise another group of substances with inhomogeneous phase state. It is just these superconductors that have mainly defined the level of application of superconducting materials in engineering. From the viewpoint of microstructure, type II superconductors are the inhomogeneous structures in which external magnetic field leads to the coexistence of the superconducting and normal phases [8]. This state of coexisting normal and superconducting phases provides high values of the critical current.

Considerable attention has been paid to investigations of FE compounds with inhomogeneous structures which are called Relaxor Ferroelectrics during recent 10-15 years. We do not refer to particular papers here because the number of publications devoted to these materials is very large, the complete understanding of the nature of their inhomogeneity has not been yet achieved, and investigation of this relaxor state is not the goal of this article.

Lead lanthanum titanate-zirconate solid solutions are the PZT-based solid solutions in which lead is partially substituted by lanthanum $(Pb,La)(Zr_{1-y}Ti_y)O_3$ (PLZT) and they also belong to above-mentioned substances [9-11]. The generalization of our recent results of studies of the crystal lattice in PLZT solid solutions with 6 at % of La and the wide interval of changes of the zirconium and titanium concentrations [12] along with the earlier published data [13,14] is presented in Fig.1- Fig.3. The "Ti-content-temperature" phase diagram (the Y-T diagram) of the solid solutions under consideration is shown in Fig.1. The dependences of the crystal lattice parameters on Ti-concentration and dependence of the fraction of rhombohedral phase in the sample volume on the Ti-content in solid solution are given in Fig.2 and Fig 3, respectively.

A distinctive feature of the phase diagram of this system of solid solutions is the existence of a large region of Ti-concentrations (the shaded region on the Y-T diagram in Fig.1) within the limits of which the state of the substance is characterized by the presence of the coexisting domains of the FE and antiferroelectric (AFE) phases. The fact that such two-phase state is more favorable energetically than the one-phase state of the FE or AFE phases is interesting from the viewpoint of physics of phase transformations. The transition of the system from the two-phase state into the single-phase FE or the single-phase AFE state can be achieved by application of the external electric field or hydrostatic pressure, respectively. The system, however, returns into the two-phase state during a certain period of time after the transition into a single phase state. The dynamics of such a spontaneous transition from single-phase state to two-phase state will be demonstrated further in this paper.

The interphase domain wall (IDW) separates the domains of the FE and AFE phases, which have different sizes of the elementary crystal cells. In connection with the above-said, two distinctive features should be emphasized. It has been shown in [15] that at the conjugation of crystal planes of the phases that have close crystal structures with different interplane distances, the intervals between dislocations are of the order of several tens of nanometers. Similar situation of conjugation of crystal planes of the phases with close crystal structures takes place in the case of the coexistence of the FE and AFE phases considered in the present paper. However, transmission electron microscopy studies of the coexisting domains of FE and AFE phases showed that linear sizes of the said domains are of order of several tens of nanometers [16, 17]. Thus, there are no dislocations at the interphase boundaries.

This signifies that the crossing of the IDW (from one phase to the other) is accompanied with continuous conjugation of the atom planes (free of breaks and dislocations). Such a coherent IDW structure leads to an increase of the elastic energy. The greater is the difference in the configuration volumes of the coexisting FE and AFE phases the greater is the significance of this increase in elastic energy. It is precisely this effect defines the positive value of the surface energy density for the boundaries separating ordinary domains in ferroelectrics [18, 19]. This elastic energy restricts the increase of the area of interphase boundaries and, consequently, the reduction in size for the domains of the coexisting phases. These stresses weaken the condition of existence of the non-uniform state.

In the substances under consideration the FE and AFE states are present. The equivalent crystallographic sites in these compounds are occupied by ions differing either in size or in charge or in both of them. In a single-phase state (or inside domains of each of the coexisting phases), each of the ions, forming the crystal lattice, is not subjected to the action of forces (more correctly, the resultant force affecting each ion is equal to zero) in the absence of external factors. Entirely opposite situation is observed for the ions located near the "bare" IDW. The balance of forces affecting each of the ions is upset. "Large" ions are pushed out into those domains, which have a larger configuration volume and, consequently, a larger distance between crystal planes. "Small" ions are pushed out into the domains with a smaller configuration volume and smaller interplane distances of the crystal lattice. Such a process is accompanied with both a decrease of the elastic energy concentrated along the "bare" IDW and an increase of the energy bound up with the segregation of the substance. The considered process of ion segregation will be completed when the new-formed IDW structure provides the energy minimum. Such "clothed" IDW will be further called real IDW or simply IDW.

Thus, the formation of the heterophase structure of the coexisting domain of the FE and AFE phases is accompanied by the emergence of chemical inhomogeneity of the substance. The said process is realized due to the diffusion of ions at relatively low temperatures ($T < T_c$). In this case the diffusion coefficients are too small, and the process is prolonged. For some PZT-based solid solutions the characteristic times of this process exceed 100 hours at 20 °C. In more details this problem will be considered in the experimental part.

Peculiarities of the solid solution local decomposition caused by the diffusion processes at relatively low temperatures due to the appearance of mechanical stresses in the vicinity of the interphase boundaries, were observed earlier [20-22]. The possibility of such decomposition was reported in [20]. In these papers we investigated inhomogeneous states with domains of coexisting FE and AFE phases in the PLLZT system $(Pb,Li_{1/2}La_{1/2})(Zr_{1-y},Ti_y)O_3$ of solid solutions. The phase Y-T diagram of PLLZT solid solutions is similar to the one for PLZT system shown in Fig.1.

The decomposition manifested itself in the form of weak diffusion lines in the X-ray diffraction patterns near the basic Debye lines characteristic of the perovskite structure of the investigated solid solutions. There were identified the dimensions of the segregates and their dependence on the location of the solid solution in the phase diagram with respect to the boundary of the regions of FE and AFE ordering. However, at that time their nature and the mechanisms of their formation were not completely understood. The shifts of the basic ions forming the skeleton of the crystal structure, in the segregates, as well as their dependence on the applied electric field were established in [21]. Finally, long-duration diffusion processes of the solid solution decomposition in the vicinity of the interphase boundaries was reported in [22]. It turns out that the duration of such a process is more than 100 hours, and it is possible to distinguish several mechanisms, which define such process.

The present paper contains the results of the first investigations of the process of local decomposition for 6/73/27 lead-lanthanum zirconate-titanate solid solution in the vicinity of the interphase boundaries, which separate the domains with FE and AFE types of dipole ordering in the bulk of the substance. This solid solution (that has the following composition: 6 at% of La, 73 at% of Zr, and 27 at% of Ti) is located in the region of the Y-T phase diagram that corresponds to such interval of solid solution compositions in which the domains of the FE and AFE phases coexist in the sample volume. X-ray analysis data shows that the rhombohedral phase is predominant one in the samples (the fraction of this phase is greater than the fraction of the tetragonal phase).

## 2. Experimental Methods

The ceramic PLZT samples were obtained by the method of joint co-precipitation from the mixture of aqueous solutions of lead and lanthanum nitrates and chlorides of titanium and zirconium with subsequent two-stage synthesis at 550 $°C$ and 850 $°C$. The polycrystalline samples were obtained by sintering at 1300 $°C$ – 1340 $°C$ in a controlled PbO atmosphere. The grain size was 6 – 8 μm. X-ray diffraction analysis confirmed that the samples were single-phase with perovskite crystal structure.

The samples' crystal structure was studied by means of a SIEMENS D-500 powder diffractometer with a Ge monochromator ($CuK_{\alpha 1}$ radiation, λ = 1.54056 Å) and a BRAUN gas position-sensitive detector. The accuracy of 2θ angle measurements was 0.01$°$.

While investigating the dynamics of the formation of segregates in the process of local decomposition of a solid solution it is necessary to measure X-ray patterns during a limited period of time within a wide interval of X-ray beam scattering angles. On the other hand, due to a low intensity of diffuse lines, the process of ageing at each point must be durable enough. Therefore, the methods of investigation of polycrystalline samples using X-ray diffractometers widely applied at present are poorly suitable for this case.

The mesoscopic structure of the solid solutions (segregation structure) was investigated by the Debye method (on the sample's metallographic section) with subsequent photometry of X-ray patterns. This method makes it possible to take measurements synchronously in a wide interval of angles, which is very important for investigation of the kinetics of the process. The $CoK_\alpha$ radiation filtered by a layer of vanadium oxide was used. The angular speed of a flat sample was 1*rpm*. The duration of the survey was 20 min (10 min for each position of the sample plane, symmetric with respect to the incident X-ray beam).

## 3. Experimental results

The mesoscopic structure of the solid solutions under investigation is caused by the coexistence of FE and AFE phase domains at temperatures below the Curie point (as shown in [21, 22] for PLLZT solid solutions) and by the local decomposition of the solid solution in the vicinity of the interphase boundaries.

The main task of the present paper is to investigate the kinetics of the formation of such structures. One can confirm this fact experimentally by subjecting the solid solution with composition from the region of the Y-T phase diagram that corresponds to the inhomogeneous structure of coexisting FE and AFE phases to quenching from high temperatures when this solution is in the PE state. The time required for the sample to establish equilibrium state with a certain value of order parameters is typically from $10^{-6}$ to $10^{-4}$ seconds, whereas the formation of segregates is caused by diffusion processes that are durable at room temperatures [22].

The samples of the solid solution under investigation, namely PLZT 6/73/27, were annealed at 600 °C for 22 hours. The diffraction patterns obtained at 600 °C (after the annealing) were characterized by the presence of singlet Debye lines only. After that the samples were quenched to room temperature. The samples were aged during a certain interval of time τ and X-ray studies were carried out. The structure of the diffraction patterns becomes more complicated in the process of the sample's ageing. Broadened diffuse lines (halos) with a significantly lower intensity appear in addition to the Debye's lines. We studied the behaviour of the halos in the two intervals of angles θ = 25° – 27° (halo 1) and θ = 29° – 32° (halo 2). The intensity, location, and shape of the halo change with time. The shape and location of the Debye's lines, which characterize the crystal structure of the solid solution under investigation, also change with time.

The dependence of the elementary cell volume (in the pseudocubic approximation) on the ageing time for the PLZT solid solution with the composition 6/73/27 is shown in Fig.4. At the first stage of the process of ageing (of about 15-20 hours) the volume of the elementary cell of the perovskite crystalline structure of the solid solution grows. At the longer ageing time the said volume decreases. As one can see, the change of the volume occurs practically according to the exponential law both at the first and the second stage of aging.

The shape, intensity and position of the halos as functions of ageing time are presented in Fig.5 and Fig.6. The following peculiarities of the behavior of the given dependences attract attention. The variation of the profile of the diffuse scattering lines takes place during the whole time of observation. The angular position of these lines also changes during the whole period of time. It should be mentioned that there is a clear correlation between the change of the positions of the Debve's lines (Fig.7) and the change of the location of the diffuse scattering lines. The dependence of the intensity of the diffuse lines on the ageing time reaches saturation in approximately 25 hours.

Changes in the shape and position of X-ray Debye's lines in the process of the samples' ageing are given in Fig.8. We analyzed the behavior of the lines (111) and (200) which are the most typical for the perovskite structure of the investigated solid solutions. The former line is a singlet at tetragonal lattice distortions, and it is a doublet at rhombohedral distortions. On the contrary, the latter line is a doublet in the first case and a singlet in the second case. Analysis of the profile of the said diffraction lines allows to make a conclusion about the change of the relation between the low-temperature phases in the process of ageing.

The profile and the location of the Debye's lines change during the whole time of the experiment as well as in the case of the diffuse scattering lines. At the first stage of the considered process, immediately after the quenching, the phase with the tetragonal type of the perovskite crystal structure distortions predominates in the bulk of the sample. Then the phase with the rhombohedral type of distortions becomes dominating. And only in the process of further ageing the equilibrium structure of the domains of the co-existing low-temperature phases is established in the bulk of the samples.

## 4. Discussion

The time dependences of the shape of diffuse X-ray lines (halos), as well as the absence of the said lines in the diffraction patterns obtained at 600 $°C$ confirm the connection between the long-term relaxation and the formation of segregates in the vicinity of the interphase FE-AFE boundaries. It is necessary to note that the establishment of equilibrium state is a long-term process in solid solutions in which the state of coexisting FE and AFE phases is realized. As one can see from the X-ray data, it continues not less then 120 hours. However, taking into account the sensitivity of this method one can assert that this process takes even longer time.

The process is a multistage one. It is clearly seen from the results given in Fig. 4-8 that there are some relaxation times caused by different mechanisms. In addition to that one has to note the mechanism responsible for establishment of equilibrium values of structural order parameters at the time intervals from $10^{-6}$ *s* to $10^{-4}$ *s* (such time intervals are beyond the abilities of our experimental methods). Without elucidating particular mechanisms for equilibrium state establishment one can however assume that the long-term character of this process is connected with the diffusion processes of the local decomposition of the solid solutions along the interphase domain boundaries. The estimation of the size of the segregates (using the shape of the diffused X-ray lines) gives the values of 8 *nm* to 15 *nm* [20, 21].

Long-term relaxation processes are non-monotonous processes due to the condition of "strong deviation from equilibrium" at the initial stage right after quenching. In the case of "weak deviation from equilibrium" (at the final stage) the relaxation process is monotonous and is described by an exponential law. Peculiarities of the crystal structure of concrete solid solution also influence the relaxation process. In particular, the PLZT and PLLZT solid solutions differ by the presence of vacancies in the A-positions of the crystal lattice of the first system of solid solutions. The relaxation process follows similar patterns for both systems. However, there is a peculiarity of this process for the second system of solid solutions at aging times 20-30 hours, which we attribute to accumulation of elastic stress and their subsequent drop [22].

Now let us dwell on the mechanisms, which define the kinetics of the processes in question at different time periods (stages). There are different mechanisms, which contribute to the long-term relaxation. Two mechanisms should be pointed out among the mechanisms that we are aware of. The contribution of the crystal lattice defects, and first of all, oxygen vacancies (under the conditions of our experiments the concentration of the vacancies in the sites of lead remained practically constant) is the first one and the second is the diffusion of the cations in the vicinity of the interphase boundaries caused by local mechanical stresses. The difference in size (and, consequently, in mobility) and charge of the ions, which are located in the equivalent sites of the crystal lattice, should be taken into account. At the same time the permanent rearrangement of the multiphase domain structure also takes place. This domain structure rearrangement happens due to the change of the local composition of the solid solution and, as a consequence, owing to the change of the local phase stability.

Not all the experimental data for the analysis of the process of long-term relaxation in the investigated PZT-based solid solutions are reported in the literature. We do not possess all the required data either. In particular, the influence of the oxygen sublattice defects on the crystalline structure of these solid solutions is investigated insufficiently. That is why in some cases we used the experimental results obtained for related oxide materials with perovskite or perovskite-type structures to clarify the nature of the observed phenomena.

Annealing of the samples at 600°C leads to the growth of the concentration of oxygen vacancies; the equilibrium concentration of the oxygen vacancies grows rapidly as the temperature rises. Quenching down to room temperature leads to freezing of the non-equilibrium elevated concentration of the vacancies in the bulk. Oxygen vacancies in ionic-covalent compositions, to which the solid solutions with perovskite structure belong, lead to increase of the crystal lattice parameters [23-25]. Alongside with it, vacancies in perovskite and perovskite-like compositions favor the increase of the stability of the phases with tetragonal type of crystal lattice distortions.

During the aging process the oxygen vacancies crop out at the sample's surface (the diffusion of oxygen into the bulk of the sample from its surface takes place) and, as a consequence, the crystal lattice parameters (registered by means of X-ray method) monotonously diminish. Since at room temperature the diffusion coefficient is comparatively low, the said process is the long-term one.

Let us now consider the processes of ion diffusion in the vicinity of the interphase FE-AFE domain boundaries. After the hardening and formation of the structure of the coexisting domains inside the sample's volume, essential mechanical stresses, caused by the difference of the interplanar distances in the neighboring FE and AFE phase domains, arise at the interphase boundaries. The growth of the elastic energy takes place. This growth gets reduced by the redistribution of the ions in the vicinity of the interphase boundary and, as a result, by the local decomposition of the solid solution and the formation of segregates. Since now mechanical stresses are the motive force of ion diffusion, this process must have a higher rate than the process of establishment of the equilibrium concentration of the oxygen vacancies. As seen from Fig.4, the dependence of the intensity of the diffuse X-ray lines, connected with the formation of the segregates, on the ageing time reaches saturation already in 20-25 hours.

The ions which, essentially differ in ionic radii (and charge) participate in the diffusion processes, which define the formation of the segregates and the local decomposition. These are lead ions $Pb^{2+}$ and lanthanum ions $La^{3+}$ which occupy the A-sites of perovskite crystal lattice, and zirconium $Zr^{4+}$ and titanium ions $Ti^{4+}$, which occupy the B-sites of the crystal lattice. Naturally, the rate of their diffusion differs as well.

It leads to the change of the chemical composition both of the segregates and the solid solutions inside the domains with time, as the ions with smaller ionic radii from the domains of one of the coexisting phases reach the interphase boundary, the solid solution inside the domains becomes enriched with 'larger' ions. As a consequence, the location of the solid solution on the phase Y-T diagram changes in time, the crystal lattice parameters increase and the type of the crystal lattice distortion changes. This is clearly seen in Fig.8. When at the first stage of the local decomposition 'small' lanthanum and titanium ions reach the interphase boundaries, the composition of the solid solution inside the domains correspond to PZT with an elevated content of zirconium. Such solid solutions are characterized by the rhombohedral type of crystal lattice distortion. Therefore, as the intensity of the diffuse lines grows, the profile of the Debye lines changes which points to the fact that the predominating amount of the tetragonal phase is replaced by that of the rhombohedral phase. That is, the share of the rhombohedrally-distorted phase in the bulk of the sample increases.

At this stage of the process the change of the elementary cell volume is defined by the competition of two processes – the reduction of the volume owing to the decrease of the concentration of the oxygen vacancies and the increase of this volume due to the enrichment of the composition inside the domains with the 'larger' ions. As a result, at the first stage the said volume increases. Its maximum value is achieved approximately in 20 hours.

As the further ageing takes place the "larger" zirconium and lead ions also reach the interphase boundaries and the composition inside the domains approaches its nominal formula composition, the solid solution regains its primary location on the Y-T phase diagram (that is, the return to the two-phase region of the state diagram takes place). In the profiles of the Debye lines this fact manifests itself (after approximately 50-60 hours of ageing) by gradual establishment of precisely such shape that is characteristic of the domains after sintering and ageing at a long time (the line shape that establishes after one year).

Simultaneously the changes in the diffusive scattering profile take place. These changes confirm the fact that the concentration of different ions in the segregates varies at the interphase boundaries. The profile of the diffusive lines is defined [21] by the resulting enveloping curve obtained after summation the X-ray scattering from the crystal planes in the segregates. Since the local chemical composition of the segregates constantly changes in the process of ageing after quenching, the profile of the diffusive lines changes as well.

The results reported in the previous paper [12] show that it is difficult enough to identify the crystalline structure of 6/100-Y/Y PLZT solid solutions at temperatures lower than that of the maximum of the ε(T) dependence in the transient region of the phase Y-T diagram. This is connected with the fact that in the bulk of the samples of those PLZT solid solutions, which belong to the shaded region in Fig.1, the domains of FE and AFE phases coexist and the effects, caused by the interaction of the domains [26-28] manifest themselves. The said coexistence of the phases also leads to certain peculiarities of the physical properties and their deviation from the properties of ordinary FE or AFE in those solid solutions, which are located in the said region of the phase diagram. These peculiarities are manifested, for example, in the dielectric or electrooptical hysteresis loops, in the dispersion of the dielectric permittivity in the vicinity of the paraelectric phase transition, in the diffuse nature of this phase transition [13].

In our opinion, additional difficulties in the identification of the crystal structure are connected with the local decomposition of the solid solutions in the vicinity of the boundaries between the coexisting phases and with the formation of the mesoscopic structure of segregates at the said boundaries in the bulk of the samples. In the X-ray diffraction patterns this manifests itself in the appearance of supplementary diffusive lines which accompany the Debye lines, on which base PLZT crystal structure is actually identified. Often these diffuse lines may appear as satellites of the Debye lines. In this case, in the process of mathematical treatment of the experimental X-ray diffraction patterns according to the existing programs a third phase is introduced into consideration. We have faced such a phenomenon while treating the experimental results. This is connected only with the fact that at the development of the corresponding programs the phenomenon of the local decomposition of the solid solution has not been taken into account. As far as we know, the effects associated with the coexistence of FE and AFE phase domains and the local decomposition of the solid solution in the vicinity of the interphase boundaries have not been discussed in the literature up to now.

At the increase of the content of lanthanum the difficulties in the identification of the structure raise. In the PLZT with 8% of La the observed picture is more complicated. The increase of lanthanum content diminishes the energy barrier that separates the free energy minima corresponding to FE and AFE states. Therefore, the interaction between these phases manifests itself greatly. The boundary region in the Y–T diagram becomes wider, and the degree of crystal lattice distortions decreases, so the X-ray splitting is less pronounced. One more important circumstance is noteworthy. The morphotropic phase boundary located at the point approximately corresponding to the composition Zr/Ti = 53/47 [29] in the Y–T diagram of $PbZr_{1-y}Ti_yO_3$, is now shifted towards the solid solutions with higher percentages of Zr as the La concentration increases. At 8–9% of La, the morphotropic boundary is observed in the vicinity of the 65/35 Zr/Ti composition [30]. This composition corresponds to the solid solutions called 'relaxor ferroelectrics' which are the object of active discussions in the literature. The substance with such composition contains three phases, and this fact is vivid in the observed structure of the X-ray lines. Each of these X-ray lines is a superposition of the ones for FE rhombohedral, FE tetragonal and AFE tetragonal phases. As far as we know, such a fact has not been considered at the identification of the crystalline structure of PLZT and, consequently, at the decomposition of X-ray diffraction lines into simple components. Here we would like to mention that the transmission electron microscopic investigations [31-34] of PLZT show that the size of the domains of the co-existing FE and AFE phases is of the order of 20–30 nm. As a concequence an additional broadening of the Debye lines is possible.

The difficulties of the identification of PLZT crystal structure are furthermore complemented by the specific features of method of hot pressing used for the preparation of the samples for which the data of crystal lattice investigations are reported in the literature. According to [35] hot pressing prevents achievement of a high degree of homogeneity due to such factors as violation of stoichiometry resulting from hot pressing, 'underannealing' effects caused by low temperatures applied in the process of hot pressing, and the presence of residual mechanical stresses arising at hot pressing. Even in the case when the hot-pressed PLZT samples have a high optical quality, some nanometer scale regions, containing chemical elements that have not reacted completely in the process of PLZT fabrication, are present in the samples' volume. In particular, this fact was confirmed in [36] by means of transmission electron microscopy. As demonstrated in [37-39], hot-pressed PLZT samples of high optical quality also contain nanodomains with the composition close to that of pure $PbZr_{1-y}Ti_yO_3$.

The results presented here and in [12] explicitly demonstrate that the crystal structure of the PLZT series with 6% of La at temperatures below the Curie temperature is complicated. One can identify it undoubtedly only if the solid solutions from different regions of the Y–T phase diagram are investigated simultaneously. There is a considerable probability of error in the structure identification when only one solid solution belonging to the boundary region of the Y–T phase diagram (shaded regions in Figure 1) is studied. It is also worth mentioning that the coexistence of FE and AFE phases or the even more complicated three-phase structure of PLZT with La content > 6% could be the reason for extraordinary behavior and unexplained phenomenon at the phase transition from the paraelectric to the dipole-ordered phase [40,41].

## 5. Conclusion

Investigation of formation of mesoscopic structures in the process of the spontaneous transition from a single-phase PE state to a two-phase state of coexisting domains of the FE and AFE phases in the solid solutions in which the coexistence of the FE and AFE phases at temperatures below the Curie point is possible has been carried out. The PLZT system of solid solutions is considered as an example. It is demonstrated that the formation of mesoscopic structures is accompanied by the local decomposition of the solid solution in the vicinity of the interphase boundaries. This process is also followed by the emergence of chemical inhomogeneity of the substance. The said process is realized by means of the ion diffusion at relatively low temperatures ($T < T_c$) when the diffusion coefficients are small, and this process is a long-time one at the room temperature, it continues not less then 120 hours. The process is a multistage one and the relaxation times at each stage are caused by different mechanisms. The size of the segregates estimated using the shape of the diffused X-ray lines has the values of 8 *nm* to 15 *nm*.

Two main mechanisms defining the kinetics of the processes responsible for formation of solid solution's mesoscopic structure at different stages are emphasized among other mechanisms. The contribution of the oxygen vacancies (in our experiments the concentration of the vacancies in the lead sites remained practically constant) is the first mechanism and the second one is the diffusion of the cations in the vicinity of the interphase boundaries caused by local mechanical stresses. The permanent rearrangement of the multiphase domain structure takes place at the same time. This domain structure rearrangement happens due to the change of the local composition of the solid solution and, as a consequence, due to the change of the local phase stability.

The local decomposition of solid solutions in the vicinity of the boundaries between the coexisting phases and the formation of the mesoscopic structure of segregates at the said boundaries in the bulk of the samples create additional difficulties in the identification of the crystal structure of the PZT-based solid solutions as well as the number of other substances that allow the coexistence of phases. This phenomenon manifests itself in the X-ray diffraction patterns by the appearance of supplementary diffusive lines which accompany the Debye lines, on which the base crystal structure of solid solutions is actually identified. These diffuse lines may often appear as satellites of the Debye lines. In this particular case a third phase is introduced into consideration in the process of mathematical treatment of the experimental X-ray diffraction patterns by means of existing methods and computer programs. This phenomenon of the local decomposition of the solid solution was not taken into account only during the development of the corresponding programs. To the best of our knowledge, the effects caused by the coexistence of the FE and AFE phase domains and the local decomposition of solid solution in the vicinity of the interphase boundaries have not been discussed in the literature up to now.

## Acknowledgments

This research was supported through the ONR Grant No. N00014-06-1-0616

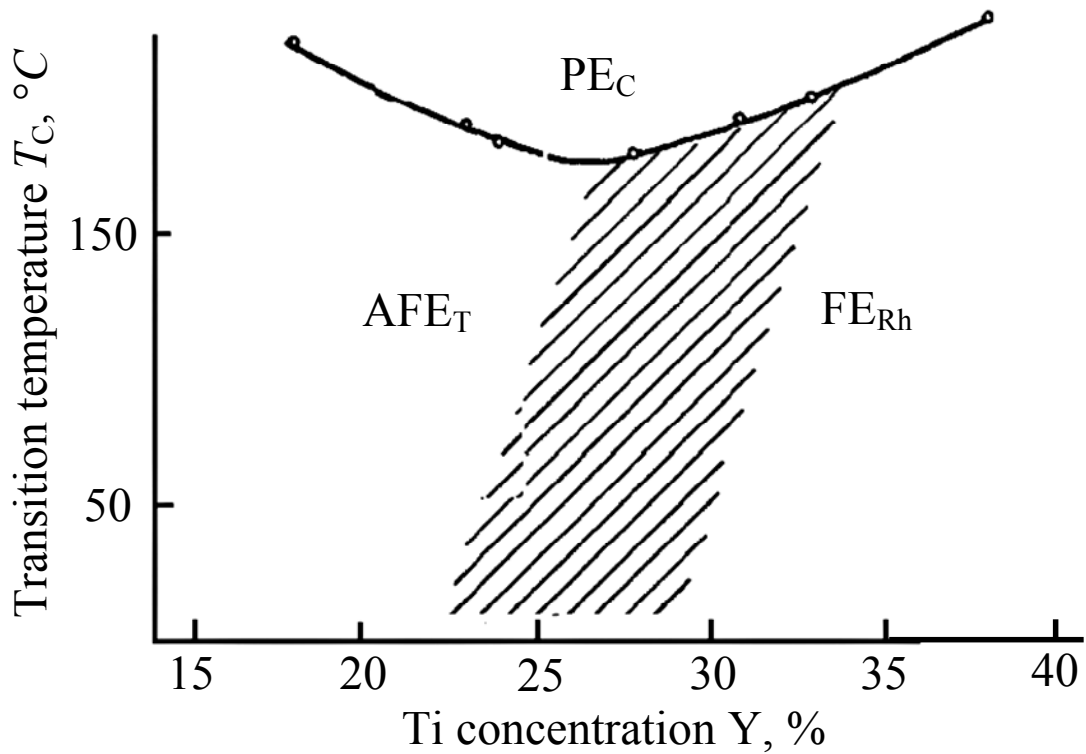

Fig.1. The "Ti-content-temperature" phase diagram for the PLZT series of solid solutions with 6% La. The region shown by dashes is the interval of the solid solution compositions characterised by the coexistence of the domain of the FE and AFE phases in the bulk of the sample.

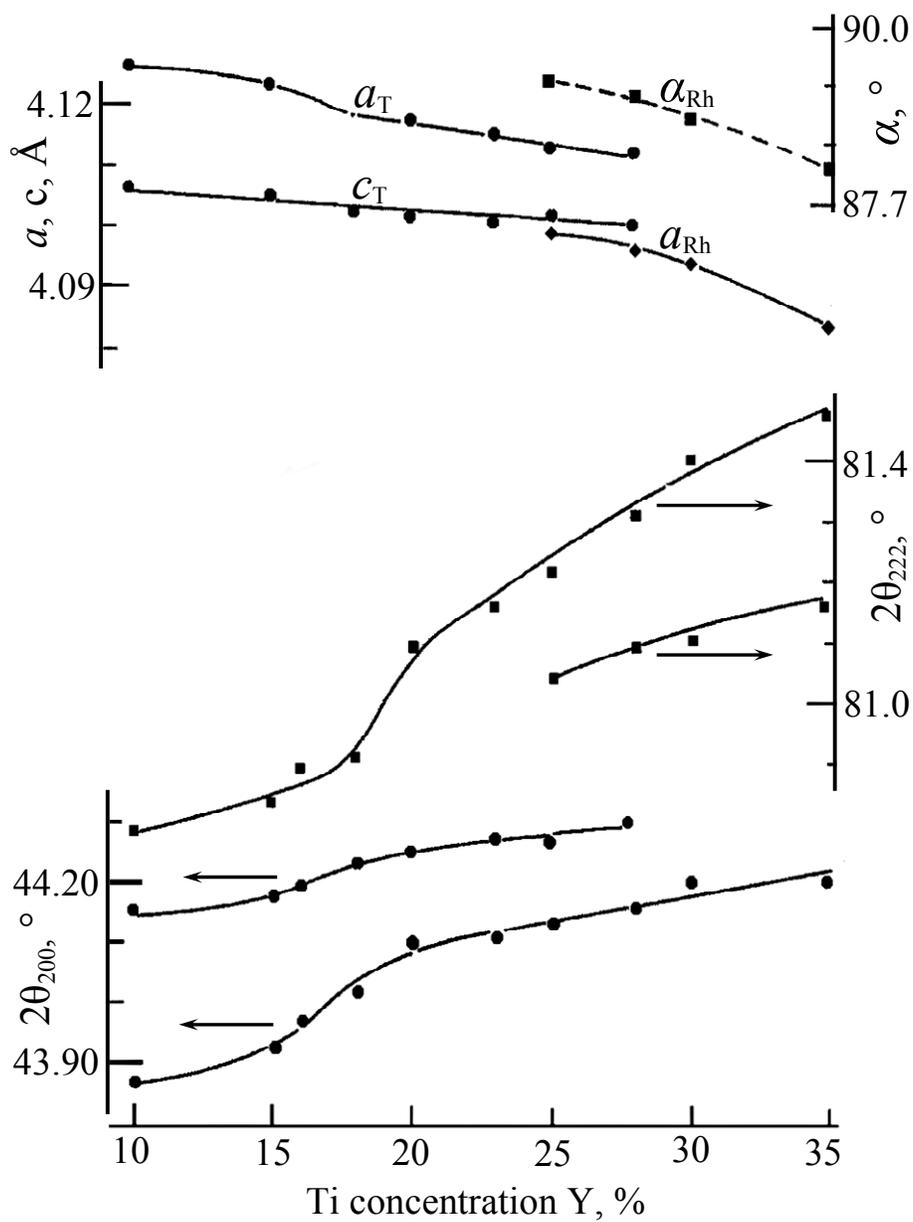

Fig.2. The dependences of the crystal cell parameters (at the top) and the positions of the peaks of components for the (200) and (222) X-ray lines (at the bottom) on Ti concentration for the PLZT series of solid solutions with 6% La.

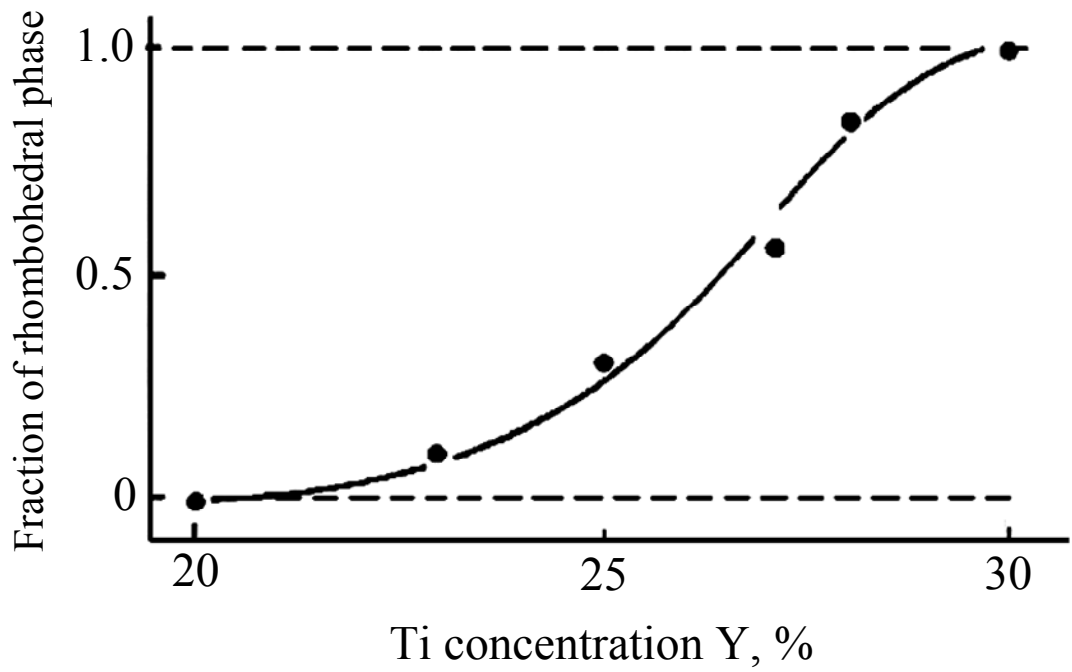

Fig.3. The dependence of the fraction of the rhombohedral phase in the sample volume on the concentration of Ti in the PLZT series of solid solutions with 6% La.

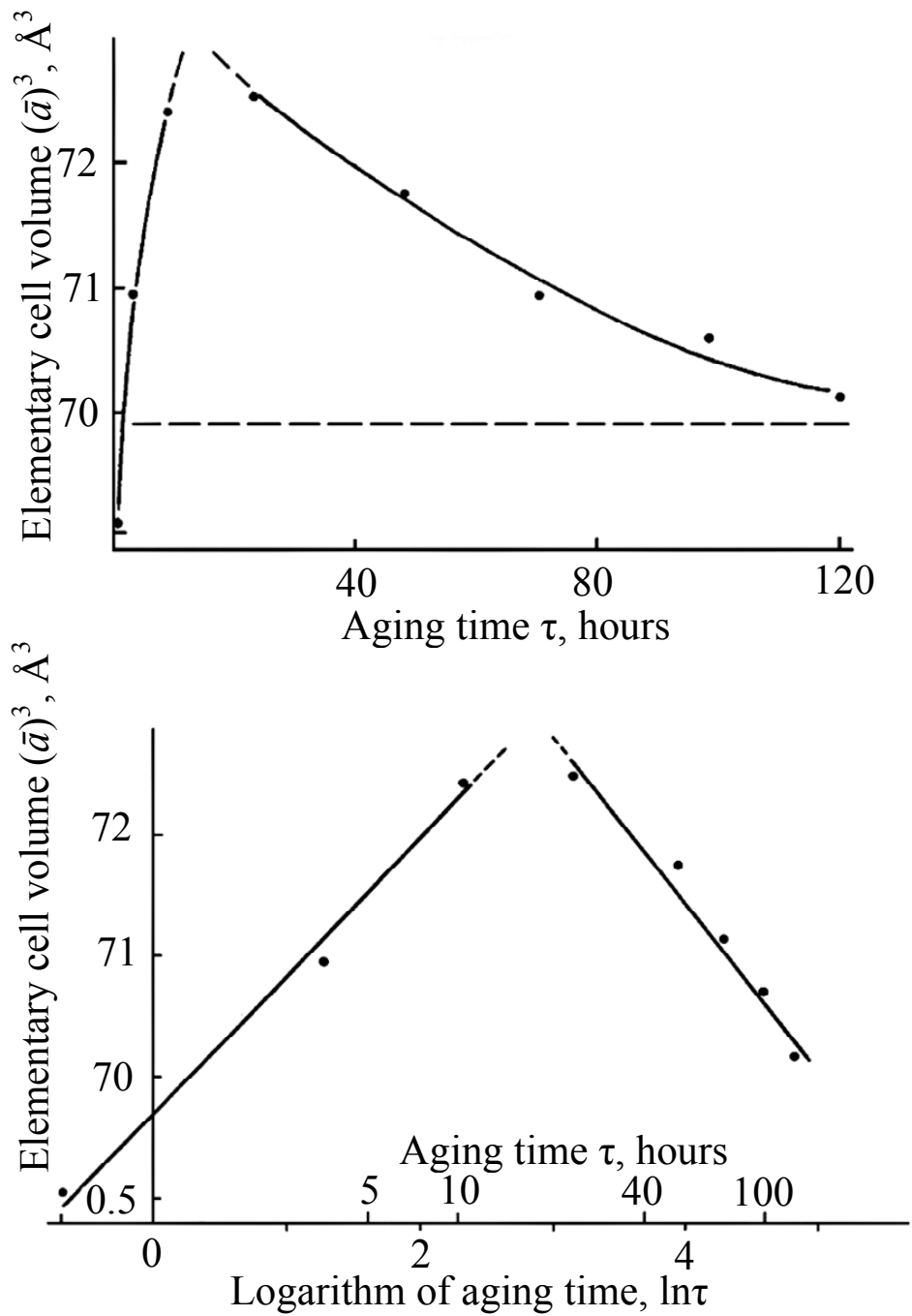

Fig.4. Dependence of the elementary cell volume on the aging time for the 6/73/27 PLZT solid solution.

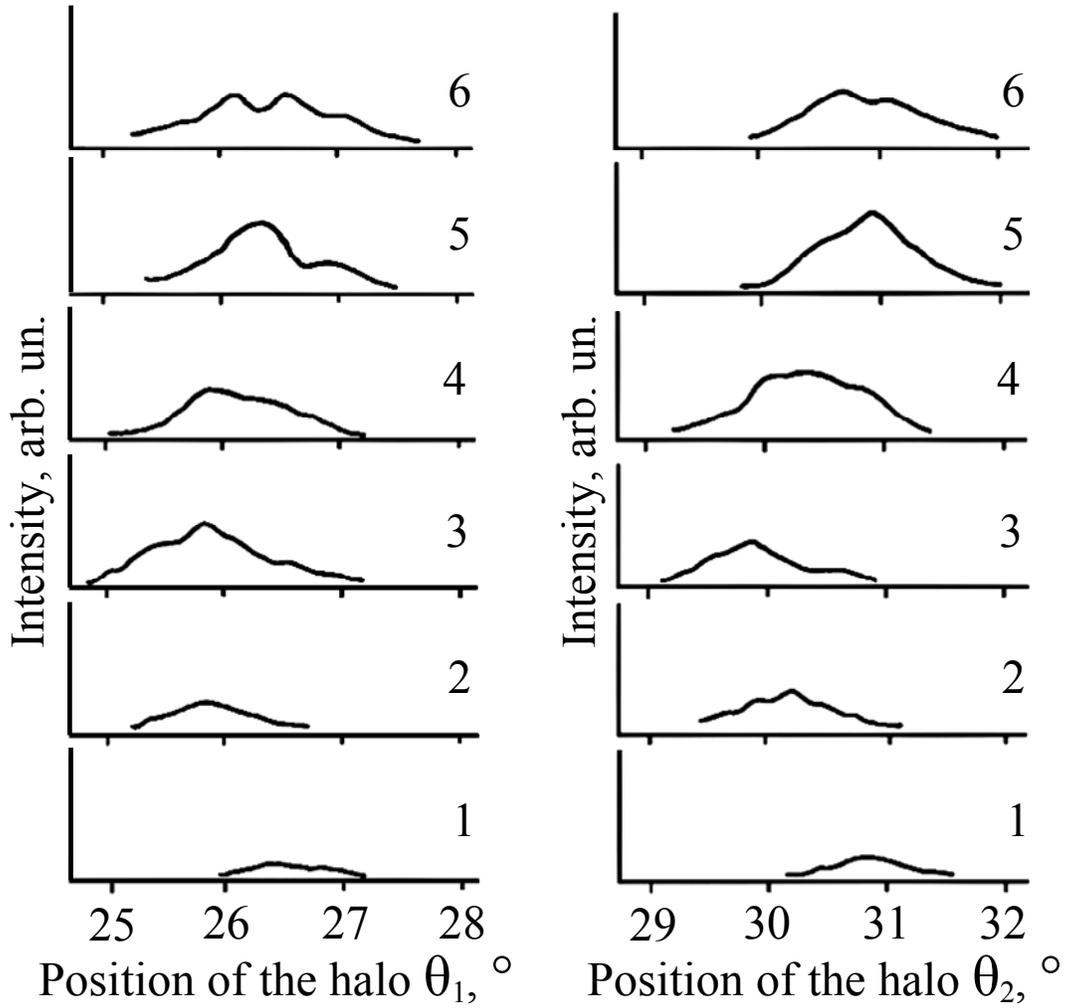

Fig.5. Changes of profile and position of the halos in the process of aging after the quenching of the 6/73/27 PLZT solid solution.

Aging time (hours): 1 – 0.5, 2 – 3.5, 3 – 23, 4 – 48, 5 – 72, 6 – 120.

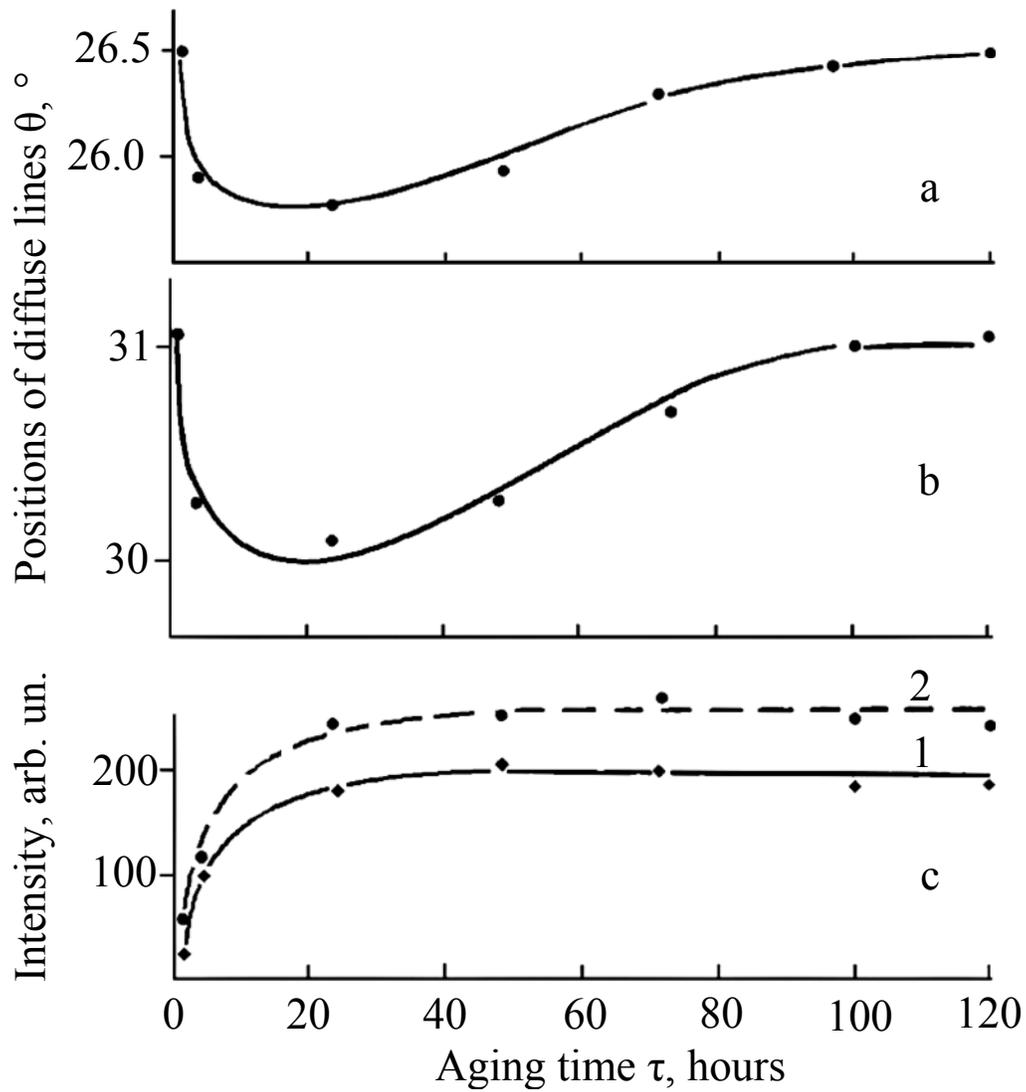

Fig.6. Aging time dependences of positions of two diffuse lines: halo 1 – a, and halo 2 – b. Dependences of the intensity of the two diffuse lines on the aging time (c): 1 – halo 1, and 2 – halo 2.

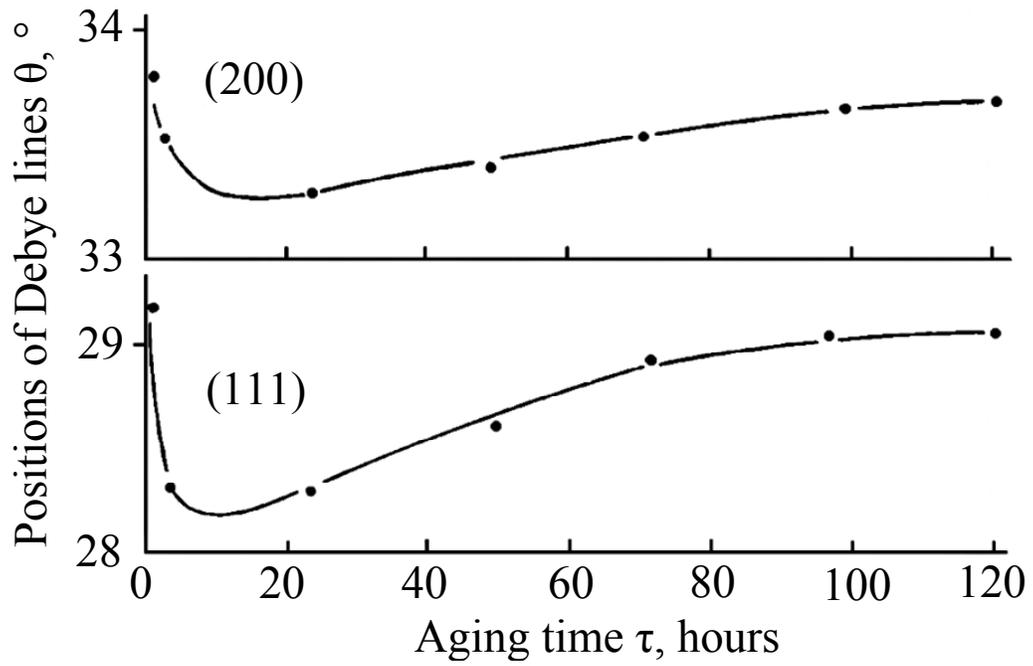

Fig.7. The dependence of positions of the (111) and (200) Debye lines on aging time for the 6/73/27 PLZT solid solution.

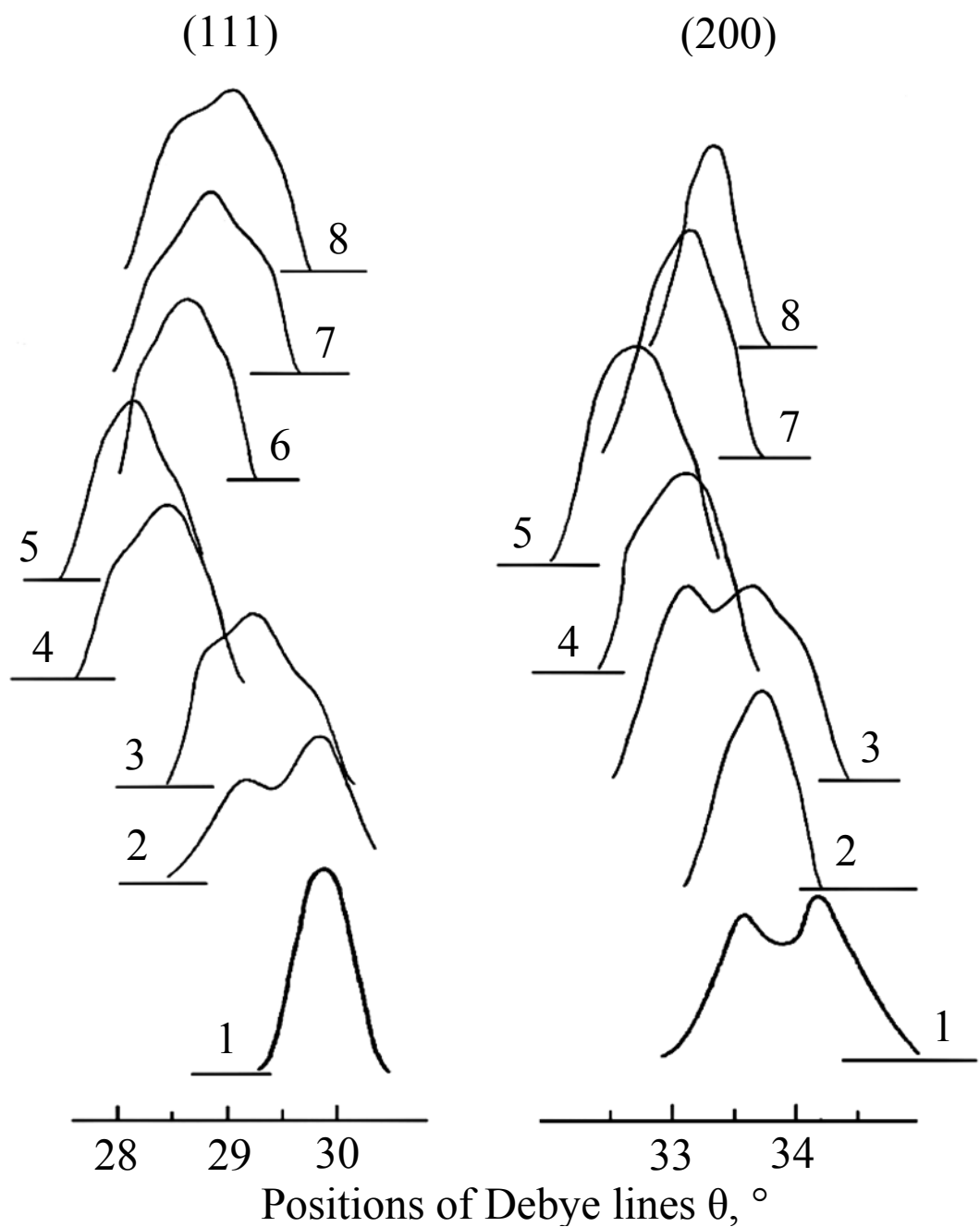

Fig.8. Variations in the shape and position of the (111) and (200) Debye lines in the process of the samples' ageing after quenching of the 6/73/27 PLZT solid solution.

Aging time (hours): 1 – 0.25, 2 – 0.5, 3 – 3.5, 4 – 23, 5 – 48, 6 – 72, 7 – 96, 8 - 120.